\documentclass[twocolumn,nofootinbib,aps,prl,showpacs]{revtex4-1}
\usepackage{dsfont}
\usepackage{graphicx}
\usepackage{float}
\usepackage{pdfpages}
\usepackage{amsmath}
\usepackage{amsfonts}
\usepackage{hyperref}
\usepackage{fullpage}
\newcommand{\mathsym}[1]{{}}

\begin{document}
\title{A new type of anomaly in turbulence}
\author{Anna Frishman$^1$ and Gregory Falkovich$^{1,2}$}
\affiliation{$^{1}$Physics of Complex Systems, Weizmann Institute of Science, Rehovot 76100 Israel \\
$^{2}$Institute for Information Transmission Problems, Moscow, 127994 Russia}
\date{\today}
\begin{abstract}
The turbulent energy flux through scales, $\bar{\epsilon}$, remains constant and non vanishing in the limit of zero viscosity, which results in the fundamental anomaly of time irreversibility.  It was considered straightforward to deduce from this the Lagrangian velocity anomaly, $\left< d u^2/dt\right>=-4 \bar{\epsilon}$ at $t=0$, where $\vec{u}$ is the velocity difference of a pair of particles, initially separated by a fixed distance. Here we demonstrate that this assumed first taking the limit $t \to 0$ and then $\nu \to 0$, while a zero-friction anomaly requires taking viscosity to zero first. We find that the limits  $t \to 0$ and $\nu \to 0$ do not commute if particles deplete/accumulate in shocks backward/forward in time on the viscous time scale. We compute analytically the resultant Lagrangian anomaly for one-dimensional Burgers turbulence and find it completely altered: $\left< d u^2/dt\right>$ has different values forward and backward in time. For incompressible flows, on the other hand, we show that the limits commute and the Lagrangian anomaly is still induced by the flux law, apparently due to a homogeneous distribution of fluid particles at all times.
\end{abstract}


 \maketitle

\textbf{\textit{Introduction}}. Flows with a little friction are very much different from those with no friction at all \cite{F}. Turbulence presented historically the first example of an anomaly, that is persistence of symmetry breaking when the symmetry breaking factor goes to zero: time reversibility is not restored at a given scale even when viscosity and the viscous scale go to zero \cite{Kolm}.
It is instructive to compare  turbulence to a quantum field theory \cite{Pol}, which stumbled upon an anomaly ten years later \cite{Schw}. There, the symmetries of the classical action can be broken by the measure of the integral over trajectories. In other words, while the classical trajectory obeys the conservation law, the average over multiple non-classical trajectories, including tunneling, does not. The symmetry is directly broken by cut-offs, introduced to regularize divergences in the measure. When the effect of the symmetry loss does not disappear as the cut-off is sent to zero or infinity, one calls it a quantum anomaly.

In turbulence, the viscosity $ \nu$ provides an ultraviolet cut-off, explicitly violating time reversibility and energy conservation. When $ \nu$ goes to zero, the range of scales where dissipation is important shrinks to zero, the Navier-Stokes equation tends to the Euler equation which is time reversible for smooth velocity fields. However, in the inviscid limit (infinite Reynolds number $Re$) the velocity becomes non-smooth and the dissipation rate $\bar\epsilon =\nu \langle|\nabla {\bf v}|^2\rangle$ has a non-zero limit equal to the energy flux through scales, $\nabla\! \cdot\! \langle {\bf u}u^2\rangle/4$, which is independent of  the scale $\delta r=|\bf r_1-\bf r_2|$ \cite{Kolm},  where ${\bf u}={\bf v}({\bf r_1})-{\bf v}(\bf r_2)$. This is called a dissipative anomaly in turbulence, since energy conservation and time reversibility remain broken in the inviscid limit.

Apart from $Re$, flows are characterized by the Mach number $M=v/c$, where $c$ is the sound speed. The anomaly above is for the incompressible limit $M\to0$. Remarkably, for weakly compressible turbulence with an effectively one dimensional flow, described by the Burgers equation, the anomaly has a similar simple form:
%(valid for perturbations with $M\ll1$ of flow past a body at $M-1\ll1$ \cite{F})
$\partial_x\langle u^3\rangle=-12\epsilon$. For multidimensional turbulence at finite $M$, the expression is more complicated \cite{A,B,C}, but the essence is the same: non-vanishing dissipation in the limit $Re\to\infty$.

Can one explain the dissipative anomaly similarly to quantum anomalies, as a symmetry breaking by a measure? One can try to do so considering fluid particles. Qualitatively, a non-smooth inviscid velocity field is non-Lipshitz, implying that even in a given velocity field particle trajectories generally are not unique. Thus, a measure corresponding to the possible trajectories should emerge somewhat analogous to the path integral in quantum field theory. This  phenomenon is called spontaneous stochasticity, it has been shown to lead to a dissipative anomaly for fields transported by the turbulent flow \cite{BGK,FGV,CFG,ED}. Non-uniqueness of trajectories seems to be a common attribute of anomalies in quantum field theory and in turbulence.  Of course, quantum non-uniqueness by itself does not provide an anomaly.

Quantitatively, in the (Lagrangian) language of fluid particles moving according to $\dot {\bf r}={\bf v}({\bf r},t)$, the anomaly was thought to be translated as follows \cite{FGV,Hill,tetr}: the rate of the relative velocity change for pairs of particles  initially separated by a fixed distance is determined solely by the flux:
\begin{equation}
\langle du^2/dt \rangle_{t=0}=%\nabla \cdot \langle {\bf u}u^2\rangle=
-4 \bar{\epsilon}\ .\label{A0}
\end{equation}
Up to now, that was the only known exact relation demonstrating a Lagrangian velocity anomaly, valid both for incompressible Navier-Stokes and  Burgers equations. Indeed, if time reversibility was restored in the inertial range in the limit $\nu\to0$, then $\langle du^2(t)/dt \rangle=-\langle du^2(-t)/dt \rangle$ and $\langle du^2/dt \rangle=0$ at $t=0$. To the contrary, (\ref{A0}) suggests that the squared velocity difference, averaged over pairs at the same distance, decreases with a rate dependent neither on distance nor on viscosity. To appreciate this result better, recall that it was also derived for an inverse energy cascade, where  $\bar{\epsilon}<0$ and $\langle du^2/dt \rangle>0$ \cite{FGV}. In other words, the Lagrangian anomaly was perceived as bringing exactly the same information (cascade rate and direction) as the Eulerian one. However, here we note that the definition of an anomaly implies taking two limits: first $\nu \to 0$ then $t \to 0$. Alas, the opposite order of limits was taken in the derivation of  (\ref{A0}). If the two limits do not commute, the relation  (\ref{A0}) is a viscous effect; reversing the order of limits should reveal an inviscid anomaly, if it exists.

This is precisely what happens in Burgers where marker particles enter/leave shocks forward/backward in time during the viscous time scale. Since particles experience a fast change in their energy only when entering/exiting shocks, the new Lagrangian anomaly manifests itself in different rates of dissipation forward and backward in time.
Indeed, the shocks are of zero measure in the limit $\nu\to0$, so that when we mark our Lagrangian particles at $t=0$, they all lie outside the shocks. As a result, there is no shock-related energy change backward in time.

On the other hand, real fluid particles of weakly compressible flows, whose velocity $v_f$ is related to $v$ from Burgers by $v_f(x)\propto v(x\pm ct)$, experience shocks moving past them with the speed of sound. They therefore do not deplete/accumulate in shocks and there is no jump in the Lagrangian dissipation rate. Still, the limits $\nu \to0$ and $t \to0$ do not commute, and when taken in the correct order the Lagrangian anomaly disappears altogether.

\begin{figure}[!ht]
    \includegraphics[width=0.9 \linewidth]{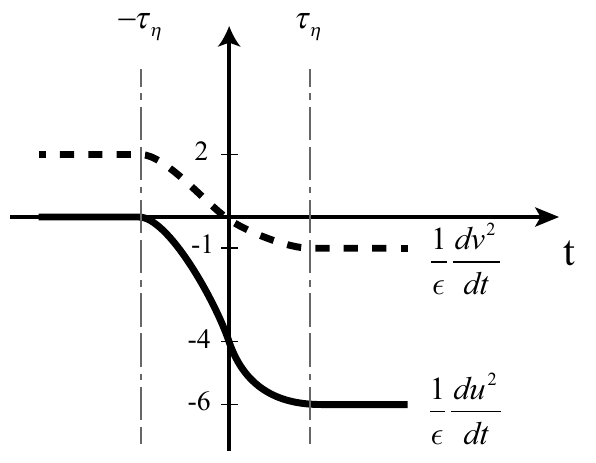}
    \centering
    \caption{A sketch of the emergence of the new Lagrangian anomaly in Burgers. The time derivative of the squared velocity of a single particle (dashed line) and the velocity difference of a pair (solid line). In the inviscid limit $\tau_{\eta}=\nu /v_{rms}^2\to 0$, the region between the vertical dashed lines shrinks, producing a discontinuity at $t=0$ which is the anomaly. The solid curve goes through $-4$ due to  (\ref{A0}).}
    \label{anomaly_figure}
\end{figure}

Below, we first present the analytic theory for Burgers turbulence. We demonstrate that the two limits do not commute. When the order of limits is taken correctly,  the nature of the anomaly is  completely different from (\ref{A0}). Backward in time the anomaly indeed vanishes in the limit of infinite Reynolds number: $\left< du^2/dt\right>\to 0$ for $t \to 0^-$. Forward in time it is larger than the  expectation (\ref{A0}):
$\left< du^2/dt\right>\to -6\bar{\epsilon}$ for $t \to 0^+$. Thus $\left< du^2/dt\right>$  has a finite jump at $t=0$  in the limit $\nu \to 0$. A similar anomaly takes place for the single particle $dv^2/dt$.
For finite $\nu$, the transition between the two values happens in a narrow time interval of order $\nu /v_{rms}^2$ - during that time the concentration of Lagrangian particles inside shocks changes. These results are summarized in Figure~\ref{anomaly_figure}.

Secondly, we will present an argument why the limits $\nu \to 0$ and $t \to 0$ do commute for incompressible turbulence, leading to the survival of the familiar Lagrangian anomaly. The same argument implies in 2d that for the direct cascade of vorticity $\omega$, a similar anomaly is also left unchanged: $d\omega/dt=\nabla\times f +\nu \Delta\omega$ gives $\langle d(\omega_1-\omega_2)^2/dt\rangle\approx-4\epsilon_\omega=-4\nu\langle |\nabla\omega|^2\rangle$.

\textbf{\emph{Burgers turbulence in 1d}}.
Consider the Burgers equation with a random force $f$ \cite{F,BK}:
\begin{equation}
\label{Bur}
\partial_t v+v\partial_x v=f+\nu \partial_{xx}v\,.
\end{equation}
It exhibits a finite energy dissipation rate in the zero-viscosity limit  due to shocks, see e.g \cite{EE,BK}.
The force correlation scale $L$ is assumed much larger than the shock width $\eta=\nu /v_{rms}$ which corresponds to the regime of dilute shocks. A Lagrangian statistical description in the inviscid limit can  be found in \cite{FGV,BB,ED}. Reiterate the often overlooked fact that $v$ which enters (\ref{Bur}) {\sl is not a fluid velocity}, it is shifted by $c t$ in space and renormalized \cite{F}. Therefore, the Lagrangian description developed in \cite{EE,BK,FGV,BB,ED} and applied below is related to markers of the Burgers velocity, not the fluid particles.

We are interested in the Lagrangian velocity moments with the initial distance between markers $\Delta$ in the inertial interval,  $\eta\ll \Delta\ll L$. The flow is assumed to be statistically homogeneous so that averages over realizations can be replaced by an average over the initial position of a pair in a given realization.
For stationary turbulence, a set of relations can be derived taking the limit $t\to0$ first \cite{F,Pol2}
\begin{equation}
\label{LKM}
\left<\frac{du^{2n}}{dt}\right>_{t=0}=-4\bar{\epsilon}_n=-\frac{2n-1}{2n+1}\rho \left<s^{2n+1}\right>\ .
\end{equation}
Here $\rho$ is the average shock density and $s$ is the shock height. For $n=1$, (\ref{LKM}) gives (\ref{A0}) and $\bar{\epsilon}_1$ is the familiar energy flux $ \bar{\epsilon}=\nu \left<(\partial_x v)^2\right>$. Let us show that upon taking the limit $\nu\to0$ first, the left equalities of the entire family  (\ref{LKM}) are replaced by true anomalies, all containing jumps, as described above for $\left<du^{2}/dt\right>$.

To explore the anomalies around $t=0$ it is enough to consider $t \ll L/v_{rms}\equiv T_L$.
At such times, the main contribution to the velocity difference $u(t)$ comes from pairs which either have a shock between them or have one marker particle inside the shock. These pairs have a velocity difference of the order of $v_{rms}$, while the rest of the pairs have $u\simeq  v_{rms}\Delta /L$.
We can thus restrict our analysis to the vicinity of a generic shock, performing a spatial average over pairs followed by an average over the parameters of the shock \cite{EE}. We begin with a single shock with prescribed parameters and calculate the spatial average which we denote by  $ \overline{u^{2n}(t)}$. Since the moments of the velocity difference are Galilean invariant, we may choose the symmetric Eulerian  velocity profile that is a standing shock:
\begin{equation}
\label{vel_eul}
v(x,t)=- w \tanh ({w x}/{2\nu} )
\end{equation}
in the segment $[-L/2,L/2]$. The shock parameters are then given by the mean shock density $\rho=1/L$ and the shock height $s=2w$, which is a random variable whose distribution is in principle derived from that of the forcing \cite{FL}.

Note that (\ref{vel_eul}) is kept stationary by an influx of particles from the boundaries while in the original problem stationarity of energy is insured by the random forcing, acting at each point. However, the forcing for the original problem has a non negligible effect on the velocity difference only for pairs that reach separations comparable to $L$. There are no such pairs at the times we are considering so the effect of the forcing can be ignored.

In the limit $\nu \to 0$,
(\ref{vel_eul}) turns into a step function with $v=w$ for $x>0$ and $v=-w$ for $x<0$. Only two groups of pairs  contribute significantly to $\overline{ u^{2n}(t)}$ in this limit: i) the pairs with the shock in between, which have $|u|=2w$, and ii) pairs with one particle inside and the other outside  the shock, which have $|u|=w$. We thus need to count the number of each type of pairs at time $t$, given that at time zero pairs were homogeneously distributed. The temporal behavior of $\overline{ u^{2n}(t)}$ can be divided into regimes belonging to three different time intervals.
At times $0<t <\Delta  / 2w$ there are $2wt$ particles inside the shock, all of which belonged to a pair with particles on both sides of the shock at $t=0$. Thus, the number of such pairs is reduced to $ \Delta-2wt$. Then, every particle  inside the shock at time $t$ belongs to two pairs, both having the second particle still outside the shock. This means there are $4wt$ pairs with one particle inside the shock. In total, for $0<t \leq \Delta  / 2w$
\begin{eqnarray}
&&\overline{ u^{2n}(t)} =(2 w)^{2n}{(\Delta -2wt) }/{L}+w^{2n}{4wt  }/{L}\nonumber\\ &&=({4w^{2n}}/{L})\left[2^{2(n-1)}\Delta-\left(2^{2n-1}-1\right)wt\right]\ . \label{u2nburg}\end{eqnarray}

For $\Delta  / 2w<t \ll T_L$ all pairs with particles on both sides of the shock have disappeared, while the particles inside the shock have their partners at a distance smaller or equal to $\Delta$ from the shock. Thus the dominant contribution comes from $2\Delta$ pairs that have $|u|=w$ which gives  $\overline{ u^{2n}(t)} =   { w^{2n}2\Delta }/{L}$.

In the third temporal segment, $t<0$, the number of pairs with a shock between them is the same as at time $t=0$ and there are no pairs with one of the particles inside the shock. Thus for $0> t \gg -T_L$
one has $ \overline{ u^{2n}(t)} =   (2w)^{2n}{\Delta }/{L}$ and   $du^{2n}(t) /{dt}=0$ for $0> t \gg -T_L$ and $ {\Delta}/{2w} < t \ll T_L$, while
\begin{align}
  \overline{ \frac{du^{2n}(t)}{ dt}}=\frac{  w^{2n+1}}{L}\left(4-2^{2n+1}\right)\ {\rm for}\  0<t < \frac{\Delta}{2w}\,.\nonumber
\end{align}

The final step is to average over $w=s/2$, denoting the probability density function for the shock height by $P(s)$, and replacing $1/L=\rho$. It gives zero for $t<0$, while for $t>0$
\begin{align}
\left< \frac{du^{2n}(t)}{dt} \right> &  =-\rho\left(1- {2^{1-2n}}\right)\int_0^{\Delta/t}\!\!\!\!s^{2n+1} P(s) ds\,. \nonumber
\end{align}
This  reveals the new type of anomaly at $t=0$:
\begin{align}
& \lim_{t \to 0^+} \left<  {du^{2n}(t)}/{dt} \right>= \rho \left(2^{1-2n} -1  \right) \left<s^{2n+1}\right>\nonumber \\
&= \left(2^{1-2n} -1  \right)  \frac{4(2n+1)}{2n-1} \bar{\epsilon}_n \,, \label{dt1}\\
& \lim_{t \to 0^-} \left<  {du^{2n}(t)}/{dt} \right>=0\,,\label{dt11}
\end{align}
different from the previously suggested (\ref{LKM}). There is no dissipation for $t<0$; the Lagrangian evolution is truly inviscid when viewed backward in time.

The footprint of this anomaly can also be seen in the two-particle conservation laws in the inviscid limit.
Following \cite{FA}, consider conservation laws of the form $\left<u^{2n} f_n(\Delta(t)/\Delta)\right>$, where $\Delta(t)$ is the separation between particles at time $t$. Backward in time, $u$  is conserved while $\Delta(t)$ grows, so $f_n(x)$ are $x$-independent for $x>1$.
Forward in time, conservation requires a power law for $x<1$, $f_n(x)\propto x^{\alpha}$ with $\alpha=-1+2^{-2n+1}$, see Appendix for the details.

The physical origin of the discrepancy between the true Lagrangian anomaly and (\ref{LKM}) lies in the change of particle distribution inside the Burgers shock. Initially it is homogeneous, but is completely altered  after a time of order $\nu/v_{rms}^2$. Backward in time, the shock is depleted of particles causing a vanishing dissipation. Forward in time, particles concentrate in the region of highest velocity gradient, where $v^2$ goes to zero at the largest rate, increasing the dissipation rate as compared with the homogeneous distribution. In the $\nu \to 0$ limit, the fraction of particles inside the shock is initially of measure zero and particles entering the shock lose $v^2$ instantaneously.

The qualitative picture described above did not rely on the properties of pairs of particles, thus we expect an anomaly of the same sort for single particles. For any time $t$ we can write
\begin{equation}
\left<  {d v^2}/{dt}\right>=2\left<f v\right>+2\nu \left<v \partial_x^2 v\right>\ .
\end{equation}
Eulerian moments are stationary and homogeneous so that forcing and dissipation cancel when taking first the limit $ t\to 0$ and then $\nu \to 0$:  $\left<f v\right>=-\nu \left<v \partial_x^2 v\right>=\epsilon$. In this case, the time derivative of the single-particle energy is zero. The anomaly, however, arises from the opposite order of limits. Reversing the order affects the dissipation but not the forcing. Forward in time, the energy loss is due to an inelastic collision between two particles with identical masses entering a shock from opposite sides. The velocity difference of the colliding particles is given by $s$ (the shock height) so that the energy lost in the process is $s^2/4$.
These collisions occur with the rate  $s/2$ per shock. Therefore, the change in $\left<v^2\right>$ due to dissipation from shocks is  $-\left< s^3/4\right>\rho=-3\epsilon$, which is naturally one half of the dissipation (\ref{dt1}) found for $\left<u^2\right>$ at $t\to 0^+$.
Combining dissipation with forcing one has $\langle dv^2/dt\rangle=-\epsilon$ for $t \to 0^+$ in the inviscid limit. At $\nu=0$ having a particle inside a shock at $t=0$ is of measure zero so that the dissipation term doesn't contribute backward in time and, due to the forcing, $\langle dv^2/dt\rangle=2\epsilon$ for $t \to 0^-$. The forward in time dissipation was indeed observed for $\tau_{\eta} <t< T_L$, in a numerical simulation of (\ref{Bur}) \cite{TA}.

The relations (\ref{dt1}) and (\ref{dt11}) imply that the limits $\nu \to 0$ and $t \to 0$ cannot be interchanged. This can also be deduced directly via a computation of $\overline{u^2}$ for the viscous Burgers equation, using the solution (\ref{vel_eul}) for a stationary shock. One then discovers a temporal dependence of the form $e^{-w^2t/\nu}$, i.e an essential singularity at $\nu=0$ from which the sensitivity to the order of limits arises. This gives a general lesson: if the limits are not interchangeable, we expect to find divergences in the next-order Lagrangian time derivatives when taking $t \to 0$ before $\nu \to 0$.  Indeed, using the Burgers equation we find
\begin{equation}
\label{disspair}
\!\!\!\!\left\langle d^2u^2/dt^2\right\rangle_{t=0}\!\approx 2\nu \left<(\partial_x v)^3\right>_{t=0}\propto 1/ \nu\,.
\end{equation}
This relation and  (\ref{A0}) are also true for particle pairs in a real compressible flow, implying that changing the order of  the limits $t\to 0$ and $\nu \to 0$ must change the anomaly (\ref{A0}). Indeed, as the fluid particles experience shocks running with the speed $c$, it is of zero measure in the limit $\nu \to 0$ for a particle to be inside a shock, both backward and forward in time. The average pair velocity difference is therefore equal to the Eulerian one at all times and the Lagrangian anomaly (\ref{A0}) disappears. For finite $\nu$, the distribution of the markers changes and stabilizes inside the shock during the time $\nu/v_{rms}c \ll \tau_{\eta}$,  so that in the interval $-\tau_{\eta} <t< \tau_{\eta}$, $\langle d u^2/dt \rangle$ goes from zero to a finite value at $t=0$ and then returns to zero.  Note also the difference between Lagrangian markers distributed uniformly at $t=0$ and the fluid density which is larger behind the shock.

\textbf{\emph{Incompressible flows.}}
Let us show that for incompressible stationary and homogeneous turbulence  there is no anomaly change as described for compressible flows. For  $\left\langle d u^2/dt \right\rangle$ to jump when $t$ passes through zero,  the second derivative $\left\langle d^2u^2/dt^2\right\rangle_{t=0}$ must diverge at $\nu\to0$ as in (\ref{disspair}). This divergence appeared in Burgers due to a product of velocity spatial derivatives at the same point, each extra derivative inside the Eulerian correlation function  bringing $1/\eta$ into the answer. To see that such a divergence is absent in an incompressible case, we follow \cite{FA} and use $d/dt=\partial_t+ v^i\nabla^i =\partial_t+\nabla^i v^i$:
\begin{equation}\langle dF(u)/dt\rangle_{t=0}=\partial_t\langle F(u) \rangle_{t=0}+ \nabla^i\langle u^iF(u)\rangle\ .\label{ME}\end{equation} The first term in the rhs is zero by stationarity.
In the last term,  incompressibility  allows us to take the $\nabla^i$ operator outside of the correlation function. That gradient kills all the single-point terms
because of spatial homogeneity. Indeed, single-point moments are known to be time-independent for homogeneous, stationary and incompressible turbulence \cite{FGV}.
As far as different-point terms are concerned, the same gradient $\nabla^i$ acting on the (finite) correlation function brings $1/r_{12}$ rather than  $1/\eta$. We conclude that the time derivative of $F(u)=d u^2/dt=2u^jdu^j/dt$ is finite at $\nu\to0$ in an incompressible case.
That demonstrates that  $\left\langle d u^2/dt \right\rangle$ does not have a jump at $t=0$ for any $\nu$ including $\nu=0$. One can establish the full commutativity of limits by showing that the derivative with respect to $\nu$ is finite as well: $\partial_\nu\left\langle d u^2/dt \right\rangle_{t=0,\nu=0}\!\!=4 \left\langle (\nabla^i v_1^j)  (\nabla^i v_2^j)\right\rangle$. This insures that the Lagrangian anomaly really is given by (\ref{A0}) for incompressible flows.
The details of all the derivations can be found in the appendices.

Importantly, the distribution of particles in incompressible flows is homogeneous at all times rather than accumulate in dissipative structures.
It implies that for incompressible flows there is no new anomaly related to the short times it takes the particle concentration in such structures to change, like in Burgers.  In particular, this shows that although the Burgers equation properly describes many physical situations it is not appropriate even for qualitative understanding of the Lagrangian properties of incompressible turbulence.

Coming back to the analogy with quantum anomalies, we mention the interpretation in terms of conflicting symmetries: a need to sacrifice one symmetry to save another. Particles colliding in a shock cannot conserve both energy and momentum, to conserve the latter they must loose the former; similarly, to conserve electric charge one violates axial charge conservation. It would also be interesting to interpret the forward/backward in time anomalies as being due to trajectory uniqueness/non-uniqueness.  Two separate trajectories entering the same shock forward in time mean non-uniqueness and spontaneous stochasticity backward in time  \cite{ED}, resulting in the anomaly forward in time. On the other hand, two different trajectories cannot meet backward in time, which leads to conservation. This conservation is rather unique for the Burgers equation, where the velocity is a martingale backward in time \cite{ED}.

It is tempting to hypothesize that the symmetric anomaly for incompressible flows follows from the degree of non-uniqueness of trajectories being the same backward and forward in time. On the other hand, we have seen that trajectories clustering on zero-measure sets forward, but not backward, in time provide for an asymmetry in the anomaly.

We thank A. Zamolodchikov, G. Eyink, T. Grafke, A. Kapustin and D. Gross for useful discussions. The work was supported by the Adams fellowship,  and grants from BSF and the Minerva Foundation.
% with funding from the German Ministry for Education and Research.

\bibliographystyle{plain}
\newpage
\onecolumngrid
\section{Appendix}
\subsection{Burgers equation with finite viscosity}
In the main text we have worked directly in the inviscid limit showing that the relations (\ref{LKM}) are replaced by a new kind of an anomaly. Here we  keep a finite viscosity, demonstrating how the anomaly arises for $\left< du^2/dt\right>$.
For this purpose we will compute $\overline{ u^{2}(t)}$, using (\ref{vel_eul}), obtaining the expression that turns into (\ref{u2nburg}) and (\ref{u2nburg2}) (for $n=1$) as $\nu \to 0$.

From (\ref{vel_eul}) one can obtain the expression for the Lagrangian velocity of a single particle starting at $x_0$
\begin{equation}
U(x_0,t)= - w \frac{\sinh\left(\frac{wx_0}{2\nu}\right)}{\sqrt{\sinh^2\left(\frac{w x_0}{2\nu}\right)+e^{\frac{w^2t}{\nu}}}}.
\end{equation}
We have set out to calculate the spatial average of $u^2(t)=(U(x_0+\Delta,t)-U(x_0,t))^2$:
\begin{eqnarray}
&&\overline{ u^{2}(t)}=\frac{1}{L}\int_{-L/2}^{L/2} w^2 \Biggl[ \frac{\sinh\left(\frac{w(x_0+\Delta)}{2\nu}\right)}{\sqrt{\sinh^2\left(\frac{w(x_0+\Delta)}{2\nu}\right)
+e^{\frac{w^2t}{\nu}}}}
-\frac{\sinh\left(\frac{wx_0}{2\nu}\right)}{\sqrt{\sinh^2\left(\frac{w x_0}{2\nu}\right)+e^{\frac{w^2t}{\nu}}}}\Biggr]^2dx_0.\label{fullu}
\end{eqnarray}
From our calculations in the inviscid limit we expect a qualitative change in the temporal behaviour around $t=\Delta/2w$, occurring during a time of the order of $\tau_{\eta}$. However, here we are only interested in the anomaly at $t=0$, so it is sufficient to focus on times $t < \Delta/(2w)-\tau$ with $\tau>0$ and $w^2 \tau/\nu \gg 1$.
Now, in order to compute (\ref{fullu}) we can divide the integration range into two: $-L/2 <x_0<-\Delta/2$ and $-\Delta /2<x_0<L/2$.

For  $-L/2 <x_0<-\Delta /2$, we can use first $w|x_0|/ \nu >w\Delta/2\nu \gg 1$ and then $\exp\left[-w(x_0+wt)/\nu\right]>\exp\left[w^2\tau /\nu\right]\gg 1$ to write
\begin{equation}
\frac{\sinh\left(\frac{wx_0}{2\nu}\right)}{\sqrt{\sinh^2\left(\frac{w x_0}{2\nu}\right)+e^{\frac{w^2t}{\nu}}}}\approx -\frac{e^{-\frac{w }{2\nu}(x_0+wt)}}{\sqrt{e^{-\frac{w}{\nu}(x_0+wt)}+4}}= -1+O(e^{-\frac{w\tau}{\nu}}).
\end{equation}
In a similar fashion, in the range $-\Delta /2<x_0<L/2$, one has
\begin{equation}
\frac{\sinh\left(\frac{w(x_0+\Delta)}{2\nu}\right)}{\sqrt{\sinh^2\left(\frac{w (x_0+\Delta)}{2\nu}\right)+e^{\frac{w^2t}{\nu}}}}\approx \frac{e^{\frac{w }{2\nu}(x_0+\Delta-wt)}}{\sqrt{e^{\frac{w}{\nu}(x_0+\Delta-wt)}+4}} = 1+O(e^{-\frac{w^2\tau}{\nu}}).
\end{equation}
Thus, (\ref{fullu}) turns into
\begin{equation}
\begin{split}
\overline{u^2}=&\frac{1}{L} \int_{-L/2}^{-\Delta /2} w^2 \left[ \frac{\sinh\left(\frac{w(x_0+\Delta)}{2\nu}\right)}{\sqrt{\sinh^2\left(\frac{w(x_0+\Delta)}{2\nu}\right)+e^{\frac{w^2t}{\nu}}}}+1\right]^2dx_0 +\frac{1}{L}\int_{-\Delta /2}^{L/2} w^2 \left[ 1-\frac{\sinh\left(\frac{wx_0}{2\nu}\right)}{\sqrt{\sinh^2\left(\frac{w x_0}{2\nu}\right)+e^{\frac{w^2t}{\nu}}}}\right]^2dx_0.
\end{split}
\end{equation}
These integrals can be computed, using Mathematica,
\begin{equation}
\begin{split}
\frac{w^2}{L}\int  \left[ 1\pm \frac{\sinh\left(\frac{wy}{2\nu}\right)}{\sqrt{\sinh^2\left(\frac{w y}{2\nu}\right)+e^{\frac{w^2t}{\nu}}}}\right]^2dy = &\frac{2 w}{L} \left(-\frac{ \nu  \text{Arctanh}\left[ \sqrt{1-e^{\frac{-t w^2}{\nu }}} \tanh\left[\frac{w y}{2 \nu }\right]\right]}{\sqrt{1-e^{\frac{-t w^2}{\nu }}}}+w y\right. \\
& \left. \pm 2 \nu  \ln\left[\sqrt{2} \cosh\left[\frac{w y}{2 \nu }\right]+\sqrt{-1+2 e^{\frac{t w^2}{\nu }}+\cosh\left[\frac{w y}{\nu }\right]}\right]\right).
\end{split}
\end{equation}

Finally, the assumptions $\Delta\ll L$, $wL/\nu \to \infty$, $w\Delta/\nu \to \infty$ as well as the time regime we chose,
allow us to write
\begin{equation}
\overline{u^2}=\frac{4 w}{L} \left(w \Delta -\frac{2 \nu  \text{ArcTanh}\left[\sqrt{1-e^{-\frac{t w^2}{\nu }}}\right]}{\sqrt{1-e^{-\frac{t w^2}{\nu }}}}\right) + O(e^{-w^2\tau/\nu}).
\label{viscu2}
\end{equation}

The result (\ref{viscu2}) demonstrates the features we have discussed above; the time dependence of the form $e^{-\frac{t w^2}{\nu }}$ makes the limits $t \to0$ and $\nu \to 0$ non commutative.
Keeping a finite viscosity in (\ref{viscu2}) one gets $\overline{du^2/dt}=-(8 w^3)/(3 L)$ at $t=0$, which after averaging over the shock hight results in (\ref{LKM}) for $n=1$:
\begin{equation}
\label{eps}
\lim_{t\to0}\left<\frac{du^2}{dt}\right>=-\frac{\left< s^3\right>}{3}\rho.
\end{equation}
On the other hand, in the inviscid limit the relation (\ref{viscu2}) is reduced to (\ref{u2nburg}) and (\ref{u2nburg2}) for $n=1$, as expected.

\subsection{Reversing limits: Burgers vs. Navier-Stokes}
As can be seen from (\ref{viscu2}), whether or not one can use the order of limits $t \to 0$ first and $\nu \to 0$ second should be apparent already from $\left<d^2u^2/dt^2\right>$ at $t=0$, which diverges in the limit  $\nu \to 0$ for the Burgers equation.
In this section we therefore present the explicit computation of $\left<d^2u^2/dt^2\right>$ in this order of limits for incompressible flows, showing that no divergence occurs at $\nu \to 0$ and demonstrating explicitly the cancelations of the single-point terms. We will then repeat this calculation for the Burgers equation, arriving at the divergent single-point term.

The first step in the calculation for an incompressible flow gives
\begin{eqnarray}
 \left\langle \frac{d^2 u^2 }{dt^2} \right\rangle_{t=0}\!\!\!\!\!=&&
2\frac{d}{dt}\left[-\left\langle  u\! \cdot\! \nabla (p_1-p_2) \right\rangle +\left\langle  u \cdot \nu \nabla^2  u\right\rangle\right] = \frac{4 d}{dt}\left[\left\langle v_1\! \cdot\! \nabla p_2\right\rangle-\left\langle v\! \cdot\! \nabla p\right\rangle+\nu\left\langle  v\! \cdot\!  \nabla^2  v-  v_1\! \cdot\!  \nabla^2  v_2\right\rangle\right]\,.\label{fullu2}
\end{eqnarray}
In the first equality we have used that the forcing is considered to be large scale compared to $R_0$ so that the forcing difference on such scales is approximately zero. Indeed
\begin{equation}
\label{forcing}
\frac{d}{dt}\left\langle  u \cdot (f_1-f_2)\right\rangle=\frac{\partial}{\partial{R_0^i}} \left[  \left\langle u^i u \cdot (f_1-f_2) \right\rangle\right]\approx 0.
\end{equation}
For the second equality in (\ref{fullu2}) we employed parity invariance and homogeneity.

As we discussed in the main text,
\begin{equation}
\label{zeros}
\frac{d}{dt}\left\langle v \cdot \nabla p\right\rangle=\frac{d}{dt}\nu\left\langle  v \cdot  \nabla^2 v\right\rangle=0.
\end{equation}

Let us pause the calculation for a moment and understand the vanishing of the single point terms through the Navier-Stokes equation.
We begin with the dissipation term,
\begin{equation}
\label{dirdis}
\nu\frac{d}{dt}\left\langle  v \cdot  \nabla^2 v\right\rangle=\nu\left\langle  v \cdot  \frac{d}{dt}\nabla^2 v\right\rangle + \nu\left\langle  (\nabla^2 v^j) (f^j+\nu \nabla^2 v^j - \nabla^j p) \right\rangle
\end{equation}
Note that one needs to be careful when calculating $d/dt\nabla^2 v$ (or $d/dt\nabla p$) since the temporal  and spatial derivatives do not commute. To avoid this complication we shall use the definition of the material derivative, $d/dt=\partial/\partial t+v^i\nabla^i$:
\begin{eqnarray}
\label{disstime}
\nu \left\langle  v \cdot \frac{d}{dt}\nabla^2 v\right\rangle  =\nu\left\langle  v^j   \nabla^2 \frac{\partial}{\partial t}v^j\right\rangle+\nu\left\langle  v^j  v^i\nabla^i\nabla^2 v^j\right\rangle
\end{eqnarray}
Due to homogeneity and stationarity the first term in the RHS of (\ref{disstime}) is equal to zero:
\begin{equation}
\left\langle  v^j   \nabla^2 \frac{\partial}{\partial t}v^j\right\rangle=-\left\langle \nabla^i v^j   \frac{\partial}{\partial t}\nabla^i v^j\right\rangle=-\frac{1}2\frac{\partial}{\partial t}\left\langle (\nabla^i v^j)^2   \right\rangle=0
\end{equation}
Next, the pressure term in (\ref{dirdis}) can also be shown vanishing, using the incompressibility and homogeneity of the flow.
Thus, the final balance is
\begin{equation}
\label{balance}
0=\nu\frac{d}{dt}\left\langle  v \cdot  \nabla^2 v\right\rangle=\nu\left\langle  v^j  v^i\nabla^i\nabla^2 v^j\right\rangle+\nu^2\left\langle  (\nabla^2 v)^2\right\rangle +\nu\left\langle (\nabla^2 v)\cdot f\right\rangle.
\end{equation}
where we expect only the first two terms in the right-hand-side not to have a well defined $\nu \to 0$ limit and thus that the cancelations of divergences are between them.

The balance for the single-point pressure term is as follows:
\begin{equation}
\label{presd}
\frac{d}{dt}\left\langle v \cdot \nabla p\right\rangle=\left\langle v \cdot \frac{d}{dt}\nabla p\right\rangle+\left\langle (-\nabla p+f+\nu \nabla^2 v) \cdot \nabla p\right\rangle
\end{equation}
Again, we can use the definition of the material derivative to write
\begin{equation}
\label{presmat}
 \left\langle  v \cdot  \frac{d}{dt}\nabla p\right\rangle  =\left\langle  v \cdot  \nabla \frac{\partial}{\partial t}p\right\rangle+\left\langle  v \cdot  v^i \nabla^i \nabla p\right\rangle.
\end{equation}
Incompressibility and homogeneity tell us that the last two terms in the RHS of  (\ref{presd}) as well as the first term in (\ref{presmat}) are zero. In addition, for the last term in the LHS of (\ref{presmat}) we can write
\begin{equation}
\left\langle  v^j \cdot  v^i \nabla^i \nabla^j p\right\rangle=\left\langle  \nabla^i \nabla^j(v^j \cdot  v^i)  p\right\rangle=-\left\langle  (\nabla^2 p ) p\right\rangle=\left\langle  (\nabla p)^2  \right\rangle
\end{equation}
with the help of incompressibility and homogeneity.
The cancellation is therefore seen explicitly
\begin{equation}
0=\frac{d}{dt}\left\langle v \cdot \nabla p\right\rangle=\left\langle(\nabla p)^2\right\rangle-\left\langle(\nabla p)^2\right\rangle=0
\end{equation}

After this brief diversion we return to the equation (\ref{fullu2}). We now calculate the time derivatives of the two point correlation functions in this equation. This calculation is very similar to the one for single-point correlation functions we performed above.
Note that
\begin{equation}
\left\langle v_1 \cdot \nabla p_2\right\rangle=\left\langle f_1 \cdot \nabla p_2\right\rangle=\nu \left\langle \nabla^2 v_1 \cdot \nabla p_2\right\rangle=0
\label{comprzero}
\end{equation}
making use of homogeneity first and incompressibility second. Also,
\begin{equation}
\left\langle  v_1^j   \nabla^2 \frac{\partial}{\partial t}v_2^j\right\rangle=-\left\langle \nabla^i v_1^j   \frac{\partial}{\partial t}\nabla^i v_2^j\right\rangle=-\frac{1}2\frac{\partial}{\partial t}\left\langle (\nabla^i v_1^j) (\nabla^i v_2^j)  \right\rangle=0
\label{dtzero}
\end{equation}
where the second equality comes from homogeneity and parity invariance and the last one from stationarity.
Thus
\begin{equation}
\label{dissp2}
\nu\frac{d}{dt}\left\langle  v_1 \cdot  \nabla^2 v_2\right\rangle=\nu\left\langle  v_1^j  v_2^i\nabla^i\nabla^2 v_2^j\right\rangle+\nu^2\left\langle  \nabla^2 v_1 \nabla^2 v_2\right\rangle +\nu \left\langle \nabla^2 v_1\cdot f_2\right\rangle.
\end{equation}
%or, alternatively, using the Navier-Stokes equation, \ref{dtzero} and \ref{comprzero} to combine the last two terms
%\begin{equation}
%\nu\frac{d}{dt}\left\langle  v_1 \cdot  \nabla^2 v_2\right\rangle =\nu\left\langle  v_1^j  v_2^i\nabla^i\nabla^2 v_2^j\right\rangle+\nu\left\langle  (v_1^i\nabla^i v_1^j)  \nabla^2 v_2^j\right\rangle.
%\end{equation}
For the pressure term we have
\begin{equation}
\frac{d}{dt}\left\langle v \cdot \nabla p\right\rangle=-\left\langle \nabla p_1 \cdot \nabla p_2\right\rangle+\left\langle  v^i_1 v_2^j \cdot  \nabla^i\nabla^j p_2\right\rangle=-\left\langle \nabla p_1 \cdot \nabla p_2\right\rangle.
\end{equation}
employing homogeneity and incompressibility in the last line.

Finally
\begin{eqnarray}
 \left\langle \frac{d^2}{dt^2} u^2 \right\rangle_{t=0}=&&-4\left\langle \nabla p_1 \cdot \nabla p_2\right\rangle +4\nu \nabla^i \left\langle  v_1^j  v_2^i\nabla^2 v_2^j\right\rangle
 -4\nu^2\left\langle  \nabla^2 v_1 \nabla^2 v_2\right\rangle \\ \nonumber
 && -4\nu \left\langle \nabla^2 v_1\cdot f_2\right\rangle
\end{eqnarray}
As discussed in the main text, all of the above terms are finite and, apart of the first two, vanish in the limit of $\nu \to 0$:
\begin{equation}
 \left\langle \frac{d^2}{dt^2} u^2 \right\rangle_{t=0}=-4\left\langle \nabla p_1 \cdot \nabla p_2\right\rangle+4\nu\left\langle (\nabla^i v_1^j)  v_2^i\nabla^2 v_2^j\right\rangle.
\end{equation}
In a 2d incompressible flow one gets a similar result, assuming a white in time small-scale forcing, where only the first term is non vanishing.

It is also possible to generalize the argument for the absence of divergences in the second time derivative of $ \left<u^2\right>$ to all time derivatives of $ \left<u^{2n}\right>$. This is made by induction, using \ref{ME} with the assumption of a stationary flow:
\begin{equation}
\left< \frac{d^{k+1} u^{2n}}{dt^{k+1}}\right>=\nabla^i \left< u^i \frac{d^k u^{2n}}{dt^k}\right>.
\end{equation}
Then, if $\left<d^k u^{2n}/dt^k\right>$ is finite $\left< u^i d^k u^{2n}/dt^k \right>$ is also finite. Finally, it is clear that acting with $\nabla^i$ on this average would not lead to divergences as it does not introduce factors of the dissipative scale.

Let us now compare our result for $\left< d^2u^2/dt^2\right>$ to that in the Burgers equation. We can again use equation (\ref{fullu2}) with the pressure set to zero, and the additional forcing term coming from compressibility of the flow cancelling as well, due to the assumption of large scale forcing:
\begin{equation}
\left\langle(\nabla \cdot \vec{v}_1)u \cdot (f_1-f_2) \right\rangle+ \left\langle(\nabla \cdot \vec{v}_2)u \cdot (f_1-f_2) \right\rangle\approx 0
\end{equation}
So we can write,
\begin{equation}
\left<\frac{d u^2}{dt}\right>_{t=0}=-4\nu  \nabla_i \left<(v_1^i-v_2^i)\vec{v}_1 \cdot \nabla^2 \vec{v}_2\right>-4\nu \left<(\nabla\cdot \vec{v})\vec{v} \cdot \nabla^2 \vec{v}\right>
+4\nu \left<(\nabla\cdot \vec{v}_1+\nabla\cdot \vec{v}_2)\vec{v_1} \cdot \nabla^2 \vec{v_2}\right>
\end{equation}
with the last terms coming from compressibility of the velocity.

At this point it is useful to restrict the calculation to the 1d flow, where one can manipulate derivatives more easily.
In that case
\begin{equation}
\left<\frac{d u^2}{dt}\right>_{t=0}= -2\nu\partial_{xxx}\left< v_1^2  v_2\right>+2\nu\partial_{xxx}\left< v_1 v_2^2\right>-6\nu \partial_x\left<v_1 (\partial_{x_2}v_2)^2\right>-4\nu \left<(\partial_x v)v \partial_x^2 v\right>.
\end{equation}
It is clear that the first two terms vanish in the inviscid limit as $\partial_{xxx}\left< v_i^2  v_j\right>$ is finite in this limit.
Then, $\left<v_1 \nu(\partial_{x_2}v_2)^2\right>$ is also finite which means that $\nu \partial_x\left<v_1 (\partial_{x_2}v_2)^2\right>$ is finite. On the other hand,  $\nu \left<(\partial_x v)v \partial_x^2 v\right>\propto 1/\nu$ so that $\nu \to 0$
\begin{equation}
\left<\frac{d u^2}{dt}\right>_{t=0}\approx- 4\nu \left<(\partial_x v)v \partial_x^2 v\right>
\end{equation}
in the inviscid limit.
This is the term which is absent for incompressible flows and which blows up as $\nu \to 0$.

\subsection{Lagrangian Conservation Laws for The Burgers Equation}
In the previous work \cite{FA} we have conjectured that generic turbulent systems must admit the family of the integrals of motion of the form
\begin{equation}
\label{consrvd}
\langle u^n f_n(\Delta x/\Delta x_0)\rangle
\end{equation}
Here $\Delta x$ is the separation between particles at time $t$, while $\Delta x_0$ is the separation at the initial time.
The forced Burgers equation is simple enough, so that it is possible to explore this conjecture here. We work with 1d Burgers, marking particles in a pair such that $\Delta x_0=x_1-x_2>0$ for all pairs, implying also $\Delta x>0$ everywhere in our calculations.

Our main claims in \cite{FA} were regarding the form of $f_n(\Delta x/\Delta x_0)$ at large times and its role as a bridge between Lagrangian dynamics and Eulerian scaling of velocities.
Here we show that the link between the Lagrangian and the Eulerian objects for the Burgers equation is a bit more subtle. Indeed, two of the assumptions we used in our arguments for incompressible turbulence do not hold. The first one is that at large times most pairs of particles have separations $\Delta x >> \Delta x_0$. Then, at such times, $f_n$ in the correlation function (\ref{consrvd}) can be replaced by its asymptotic behaviour at infinity.
On the contrary, for the Burgers equation at asymptotically large  times most pairs will have zero separation - they would have dived into a shock. The remaining pairs, those that contribute to the average, spread over a range of separations.

The second assumption was that the initial separation $\Delta x_0$ is forgotten at large times. This is not the case for Burgers, where the Lagrangian moments of velocity depend on $\Delta x_0$ at all times. Indeed, the contribution of pairs to the average, at any time, depends upon the presence of a shock between them, the probability for which scales like $\Delta x_0$.
Thus here $f_n$ does not play the role of a link to the Eulerian velocity scaling. This scaling is always provided by the presence of the shock.

None the less, one can still speak of Lagrangian conservation laws for the Burgers equation. The idea is to balance  the change in time of the velocity by using the function $f_n(\Delta x/\Delta x_0)$.
We have already seen in the main text that there exist three qualitatively different time regimes for $u^{2n}$, each having it's own set of  pairs that provide the main contribution to $u^{2n}$. Let us briefly review what these are, including the dependence of their separations and velocities on the parameters in the problem.

\begin{itemize}
\item Times $-T_L\ll t<0$:
The pairs that contribute for such times are those with a shock between the particles. Each pair has the velocity difference $u=-2w$ and the separation between particles equal to $\Delta x=\Delta x_0-2wt>\Delta x_0$. The probability to have such a pair is $ \Delta x_0 \rho$.
\item Times $0<t \leq \Delta x_0/2w$:
Pairs with a shock between particles, and thus velocity difference $u=-2w$, have the separation between particles equal to $\Delta x=\Delta x_0 -2wt$. The number of such pairs per shock is $\Delta x_0 -2wt$. Then, there are also pairs with one particle in the shock and one outside, which have $u=- w$. Particles in such pairs either used to be on the same side of the shock, in which case they have separations in the range $\Delta x_0-wt\leq \Delta x\leq\Delta x_0$, or, if they began on different sides of the shock, then $\Delta x_0 -2wt\leq\Delta x\leq\Delta x_0-wt$.
There are two pairs per shock for each such separation, the particle outside the shock being on one or the other side of the shock.

We can summarize these results by saying that in this time interval there is a probability of $2\rho$ to have a pair of particles with velocity difference $u=-w$ and separation in the range $\Delta x_0 -2wt\leq\Delta x\leq\Delta x_0$. With probability $\rho (\Delta x_0 -2wt)$ a pair would have
$u=-2w$ and separation $\Delta x=\Delta x_0 -2wt$. Other pairs have negligible velocity differences and do not contribute (\ref{consrvd}).

\item Times $ \Delta x_0  / 2w\leq t \ll T_L$: Only pairs with one particle inside the shock have a non negligible velocity difference between particles. That velocity difference is $u=-w$ and the particle separation is in the range $0<\Delta x<\Delta x_0$. The probability to have a pair of this type is $2\rho$.

\end{itemize}
Now, to compute $\overline{u^{2n}f_n(\Delta x/\Delta x_0)}$ at a specific time $t$, all we need to do is to average over the separations and velocity differences of the pairs listed above, in the relevant time interval, using their respective probabilities.
Here we are interested in finding $f_n(\Delta x/\Delta x_0)$ such that this average is time independent.

First we notice that pairs have $\Delta x>\Delta x_0$ only backward in time. Thus we can find $f_n(z)$ for $z>1$ based purely on considerations for times $-T_L\ll t<0$. As the velocity difference is conserved backward in time, it is enough to choose  $f_n(z)=f_n(1)$ for $z>1$ so that in the range $-T_L\ll t<0$
\begin{equation}
\overline{u^{2n}f_n(\Delta x/\Delta x_0)}=\rho \Delta x_0 (2w)^{2n} f_n(1)
\end{equation}
is time independent.

The time range $ \Delta x_0  / 2w\leq t \ll T_L$ is also quite simple to treat. For these times
\begin{equation}
\begin{split}
\overline{u^{2n}f_n\left(\frac{\Delta x}{\Delta x_0}\right)}=2\rho w^{2n} \int_0^{\Delta x_0}f_n\left(\frac{\Delta x}{\Delta x_0}\right)d(\Delta x)
\end{split}
\end{equation}
which is manifestly time independent. This result, of course, has to match with that at $t=0$ (or any other time), which occurs by construction from the demand of a conservation law at times $0<t \leq \Delta x_0/2w$, to which we turn next.

We have that
\begin{equation}
\overline{u^{2n}f_n\left(\frac{\Delta x}{\Delta x_0}\right)}=\rho(\Delta x_0-2wt)(2w)^{2n}f_n\left(\frac{\Delta x_0-2wt}{\Delta x_0}\right)+ 2\rho w^{2n}\int_{\Delta x_0-2wt}^{\Delta x_0} f_n\left(\frac{\Delta x}{\Delta x_0}\right)d(\Delta x)
\label{fnav}
\end{equation}
for the times $0<t \leq \Delta x_0/2w$.
Requiring the time derivative of (\ref{fnav}) to be zero and replacing $z=\frac{\Delta x_0-2wt}{\Delta x_0}$ produces the following equation
\begin{equation}
zf_n'\left(z\right)=f_n\left(z\right) \frac{\left(4-2^{2n+1}\right)}{  2^{2n+1} }\,,
\end{equation}
which has the solution
\begin{equation}
f_n(z)=f_n(1)z^{-1+\frac{1}{2^{2n-1}}}.
\end{equation}

As promised, one can check that the result for $t>\frac{\Delta x_0}{2w}$ matches that at $t\leq 0$
\begin{equation}
\begin{split}
\overline{u^{2n}f_n\left(\frac{\Delta x}{\Delta x_0}\right)}= 2 \rho w^{2n} \int_0^{\Delta x_0}\!\!\!\!f_n\left(\frac{\Delta x}{\Delta x_0}\right)d(\Delta x)=2\rho \Delta x_0 w^{2n}\int_0^{1}\!\!\!\!f_n(1) z^{-1+2^{-2n+1}}dz= \rho \Delta x_0 (2w)^{2n}f_n(1).
\end{split}
\end{equation}

Combining the backward with the forward in time results for the conservation law we find
\begin{align}
& f_n(z)=f(1)z^{-1+2^{-2n+1}} && z \leq 1 \nonumber \\
\label{fncon}
& f_n(z)=f(1) && z\geq 1.
\end{align}
It is clear that since $\overline{u^{2n}f_n\left(\frac{\Delta x}{\Delta x_0}\right)}$ with (\ref{fncon})
is time independent it will remain time independent also after the average over the shock parameters. We have thus found the following conservation law forward in time
\begin{equation}
\left<\frac{u^{2n} }{(\Delta x)^{1-2^{-2n+1}}}\right>=\rho \left< s^{2n}\right> \Delta x_0^{2^{-2n+1}}
\end{equation}
while backward in time the conservation law is the velocity itself $\left<u^{2n}\right>=\rho \left< s^{2n}\right> \Delta x_0$.

For example for $n=1$, forward in time
\begin{equation}
\left<\frac{u^{2} }{\sqrt{\Delta x}}\right>=\rho \left< s^{2n}\right> \sqrt{\Delta x_0}
\end{equation}
and for $n\to \infty$ we have
\begin{equation}
\lim_{n\to \infty}\left<\frac{u^{2n} }{(\Delta x)^{1-2^{-2n+1}}}\right>\to \lim_{n\to \infty}\left<\frac{u^{2n} }{\Delta x}\right>\to \rho\lim_{n\to \infty} \left< s^{2n}\right>
\end{equation}
Of course, all the calculations above are correct only in the inviscid limit, for $\Delta x_0 \ll L$ in a stationary flow.

\end{document}